\documentclass[preprint,showpacs,prb,aps,amsmath]{revtex4}
\def\etal{{ \it et al. }}
\def\PRB{Phys. Rev. B}
\usepackage{graphics}
\usepackage{graphicx}
\usepackage{natbib}

\begin{document}
\title{ Superconductivity in Boron under pressure - why are the measured T$_c$'s so low? }
\author{S. K. Bose}
\email[
]{sbose@brocku.ca}
\affiliation{Department of Physics, Brock University, St.\ Catharines, Ontario
L2S 3A1, Canada and \\Max-Planck-Institute for Solid State Research,
Heisenbergstr.\ 1, D-70569 Stuttgart, Germany 
}
\author{T. Kato}      
\affiliation{
Institute for Fundamental Chemistry
   34-4, Takano-Nishihiraki-cho,
	 Sakyo-ku, Kyoto 606-8103, Japan and\\
	 Max-Planck-Institute for Solid State Research,
	 Heisenbergstr.\ 1, D-70569 Stuttgart, Germany}

\author{O. Jepsen}
\affiliation{Max-Planck-Institute for Solid State Research,
Heisenbergstr.\ 1, D-70569 Stuttgart, Germany}
\date{\today}

\begin{abstract} 
Using the  full potential linear muffin-tin orbitals (FP-LMTO) method we examine the  
pressure-dependence of superconductivity in the two metallic phases of Boron: bct and fcc. 
Linear response calculations are carried out to examine the phonon frequencies and electron-phonon
coupling  for various lattice parameters, and superconducting transition temperatures are
obtained from the isotropic Eliashberg equation.
The fcc phase is found to be stable only at very high pressure (volume per atom $<$ 21.3 bohr$^3$), 
estimated to be in excess of 360 GPa. The bct phase (volume per atom $>$ 21.3  bohr$^3$) is 
stable at lower pressures
in the range of 210-360 GPa. In both bct and fcc phases the superconducting transition temperature
T$_c$ is found
to decrease with increasing pressure, due to stiffening of phonons  with an accompanying
decrease in electron-phonon coupling. This is in contrast to a recent report, where T$_c$ is found to
increase with pressure. Even more drastic is the difference between the measured T$_c$, in the range 4-11 K,
and the calculated values for both bct and fcc phases, in the range 60-100 K. The calculation 
reveals that the transition from the fcc to bct phase, as a result of increasing volume or decreasing
pressure, is caused by the softening of the X-point transverse phonons. This phonon softening also
causes large electron-phonon coupling for high volumes in the fcc phase, resulting in 
coupling constants in excess of 2.5 and T$_c$ nearing 100 K.

Although it is possible that the method used somewhat overestimates the electron-phonon coupling,
its success in studying several other systems, including MgB$_2$, clearly suggests that 
the experimental work
should be reinvestigated. We discuss possible causes as to why the experiment might have revealed
T$_c$'s much lower than what is suggested by  the present study. The main assertion of this
paper is that the possibility of high T$_c$, in excess of 50 K,
in high pressure  pure metallic phases of boron cannot be ruled out,
 thus pointing to (substantiating)  the need for further experimental 
 investigations  of the superconducting properties
of high pressure pure phases of boron. 
\end{abstract} 


\maketitle 
\section {INTRODUCTION} 
Although metallization of boron under pressure was predicted on the basis of electronic 
structure calculations some time ago \cite{mcmahan} , experimental verification of this result 
has so far been lacking. On the basis of the calculations \cite{mcmahan} nonmetallic icosahedral
boron(B) is expected to undergo a structural transition, first to a body-centered tetragonal (bct)
phase at $\sim$210 GPa and then to a face-centered cubic (fcc) phase at $\sim$360 GPa. Recently
high pressure experiments by  Eremets \etal \cite{science} have found boron to be not only a metal, but
also  a superconductor at high pressure, with superconducting transition temperature $T_c$ increasing
with increasing pressure. Superconductivity appears at around 160 GPa, and $T_c$ increases from 6 K
at 175 GPa to 11.2 K at 250 GPa. The appearance of superconductivity in B under pressure is not
surprising, because the existence of different phases is usually an indication of strong 
electron-phonon coupling, which is  (most often) also responsible for superconductivity.  
In this work we study the superconductivity in  bct and fcc phases of B using {\it ab initio} theoretical
calculations. We use the linear response scheme implemented by Savrasov \cite{savrasov1,savrasov2}
to study the phonon properties and electron-phonon coupling. The electronic structure is calculated
by using the full potential linear muffin-tin orbital (FP-LMTO) method \cite{savrasov-el}.
Recently such calculations have been performed  for MgB$_2$\cite{kong} and several
other systems\cite{cacuo3,bose-hcp} including some boron containing compounds\cite{libc,unpub}, 
and have been found highly
successful in capturing the essence of electron-phonon superconductivity in these compounds.

A theoretical study of superconductivity in the fcc phase of boron was reported by
Papaconstantopoulos and Mehl\cite{papa} shortly after Eremets \etal \cite{science} reported their
experimental results. However, the lattice parameters for the
fcc phase used by the authors correspond to volumes per atom for which a lower energy structure is bct,
as had been revealed by an earlier calculation of  Mailhoit \etal \cite{mcmahan}.
 Papaconstantopoulos and Mehl\cite{papa}  calculate the electron-phonon
coupling or the Hopfield parameter  using the rigid muffin-tin approximation,
arguing that the fcc phase might be metastable for such lattice parameters. This speculation is based
on their calculation of elastic constants showing $C_{11}-C_{12} > 0; C_{12} > 0$, i.e., 
the Born criteria\cite{born}
for the stability of  the cubic phase are not violated. 
However, our linear response calculations yield imaginary 
 phonon frequencies  for lattice parameters in excess of 4.4 a.u. in the fcc phase.
 Thus, although the existence of a metastable fcc phase for lattice parameters 4.6-6.0 a.u. used by
 Papaconstantopoulos and Mehl\cite{papa} cannot be ruled out, the chances are that the equilibrium phase
 at such lattice parameters is not fcc. Our calculations for equilibrium fcc phase 
 (lattice parameter less than
 4.4 a.u.) shows a volume dependence of the superconducting transition temperature that is 
 the opposite of what
 is observed experimentally, i.e., $T_c$ decreases with increasing pressure. 
  The study by Papaconstantopoulos and Mehl\cite{papa} only involves the fcc phase with large lattice
  parameters, for which the structure is unstable or, at best, metastable. The present work studies
  both the fcc and bct phases, each within its appropriate range of lattice parameters. 
Our linear response calculations also indicate the onset of a bct phase via preference for a change
in the $c/a$ ratio as the volume per atom is increased. Detailed discussion of the relevance of our results to
the experimental work and that of Papaconstantopoulos and Mehl\cite{papa} is provided in appropriate sections.

The organization of this paper is as follows. In section II we discuss the stability of the bct and fcc 
phases  of boron as a function
of volume per atom. In Sections III and IV we discuss 
the calculated electronic bands, phonon spectra, and electron-phonon
coupling for the fcc and bct phases, respectively. 
In section V we discuss the volume-dependence of the superconducting transition temperature, 
  summarize and compare our results with the experimental data and discuss 
 possible sources of difference between the two.

\section{Stability of fcc and bct phases}

\begin{figure}
\includegraphics[angle=270,width=12 cm]{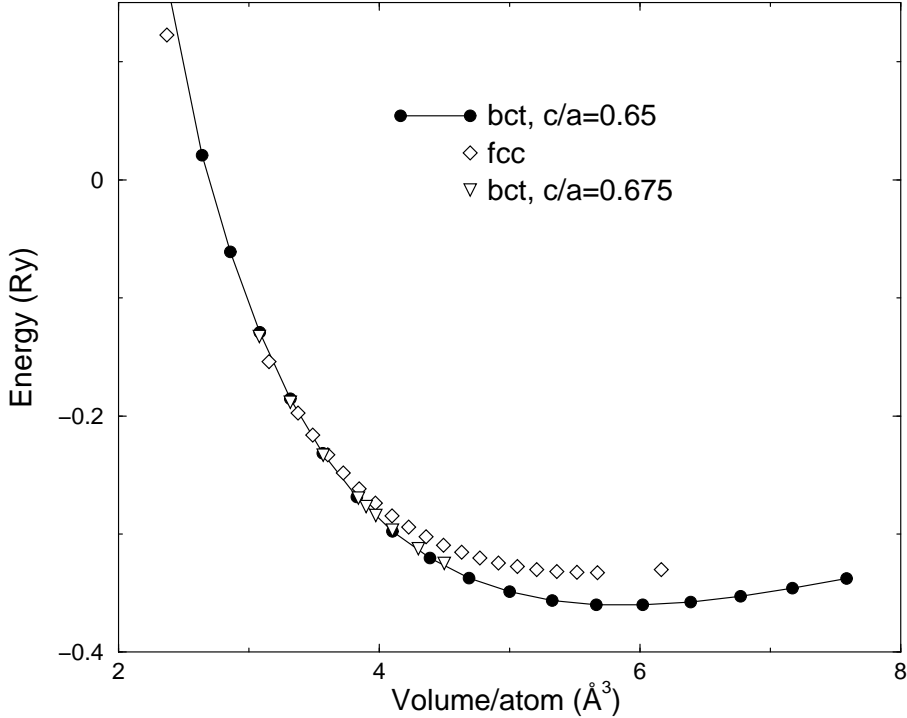}
\caption{Energy vs. volume (per atom) curves for the fcc and bct phases of boron.}
\label{fig1}
\end{figure}

The volume vs. energy curves for fcc and bct phases of boron, obtained via  FP-LMTO calculations, are
shown in Fig.\ref{fig1}.  A constant energy of 49.0 Ry. has been added to the energies per atom
for convenience in plotting. These results are very similar to those given by LMTO-ASA method, 
and are in close agreement with the earlier  plane-wave pseudopotential calculations of 
Mailhoit \etal \cite{mcmahan}.
 Pseudopotential calculations of total energy of the bct structure by Mailhoit \etal \cite{mcmahan}
 as a function of $c/a$  had indicated a global minimum around $c/a\sim 0.6$.  Via an elaborate study
 of the total energies   for various  monoclinic and tetragonal distortions of the fcc unit cell
  as well as the bct total energies for various $c/a$ values
 these authors concluded that
 the bct minimum occurs around $c/a\sim 0.65$. We have not carried out extensive calculations to locate the
 exact value of $c/a$, but our FP-LMTO calculations on a crude mesh of $c/a$ for fixed volumes
 do indicate a value in the range
   $0.6-0.7$. Our linear response calculations for $c/a = 0.65$, however,
   produced complex phonon frequencies at the symmetry point
 N and nearby wave vectors. Increasing $c/a$ to $0.675$ resulted in real frequencies 
 at all symmetry and intermediate points. Thus the 
 linear response results in this paper are presented for this value of $c/a$. It is evident 
 from Fig.\ref{fig1}
 that the actual energy values for $c/a=0.65$ and $0.675$ for the same volume in the 
 bct phase are very close, and 
  their energy-volume curve intersects the fcc energy-volume curve at almost the same point. 
  Thus the conclusions regarding the stability of the
 bct phase against fcc are not influenced critically by the choice of $c/a$ in the range $0.65-0.675$.

 \section{The fcc Phase}
 \subsection{Energy bands}
\begin{figure*}
\includegraphics[angle=270,width=8.0cm]{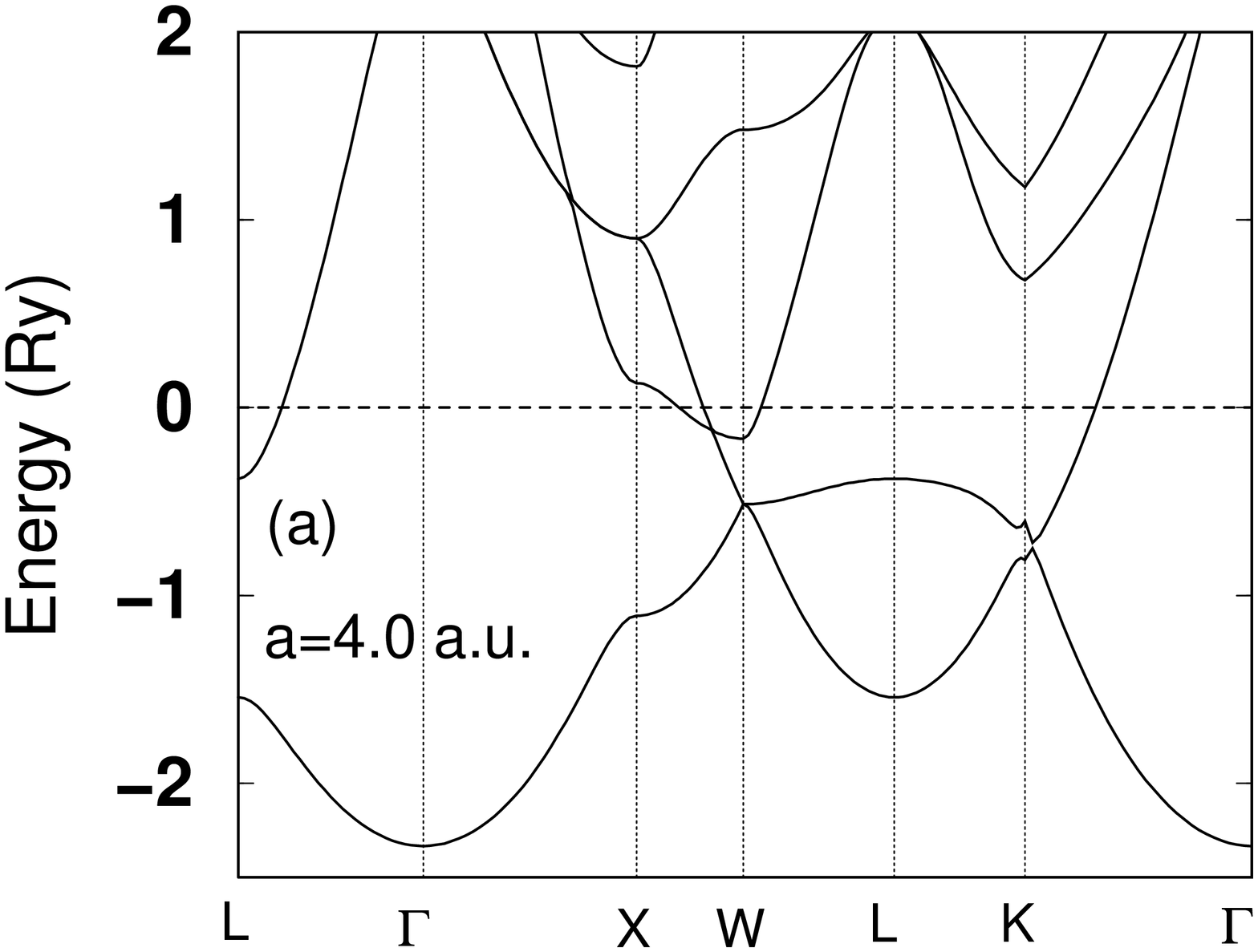}
\includegraphics[angle=270,width=8.0cm]{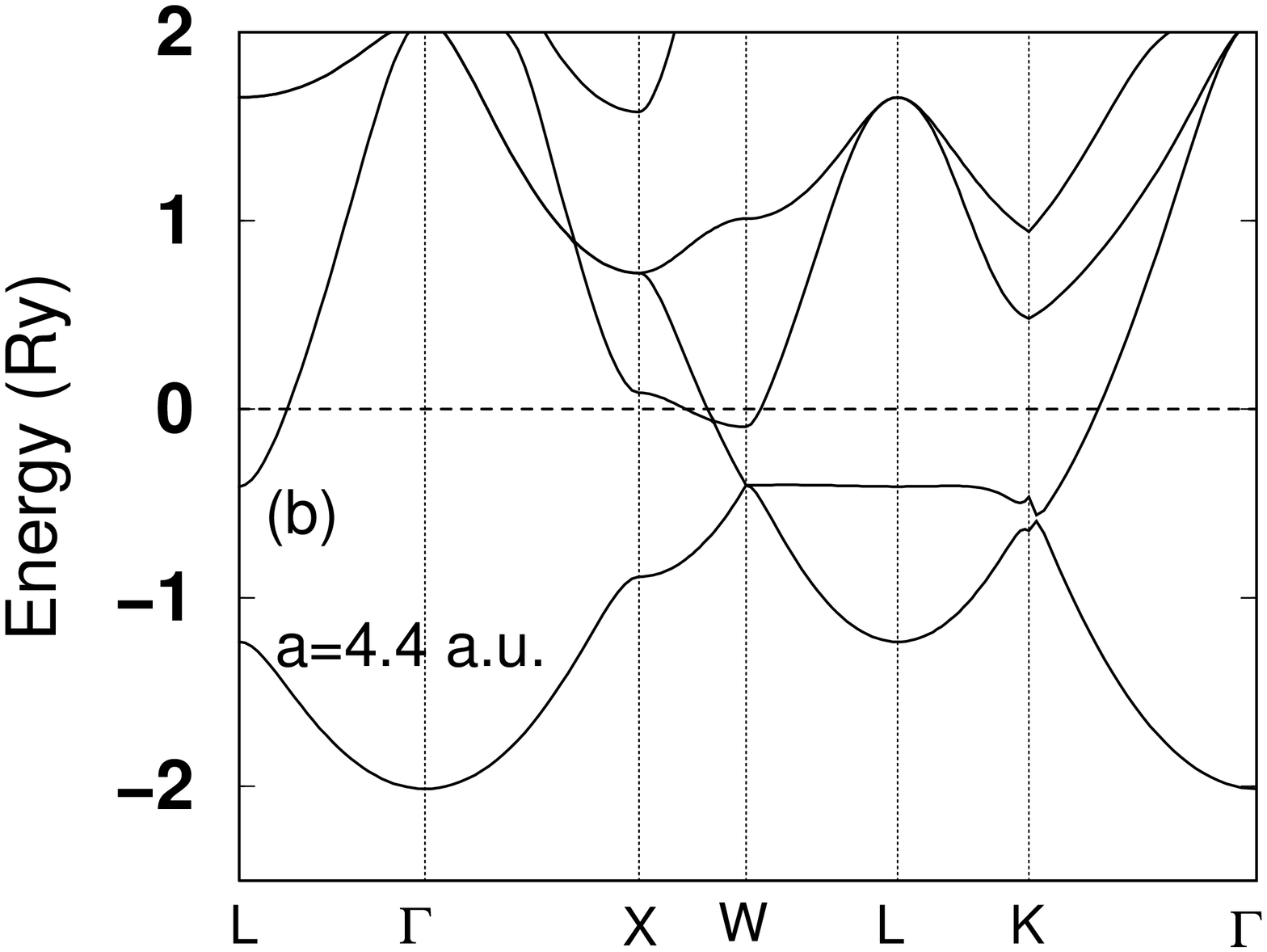}
\includegraphics[angle=270,width=8.0cm]{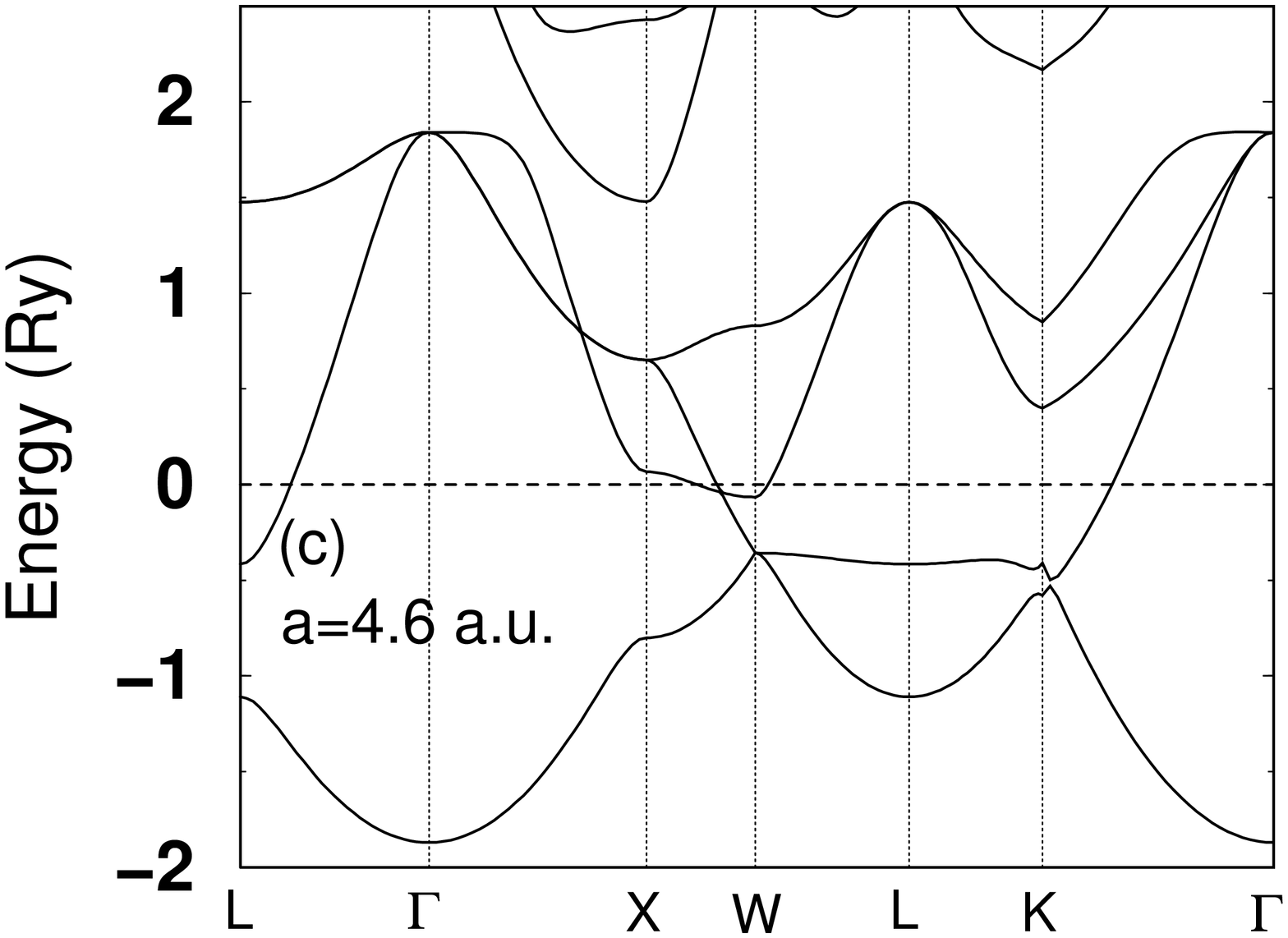}
\includegraphics[angle=270,width=8.0cm]{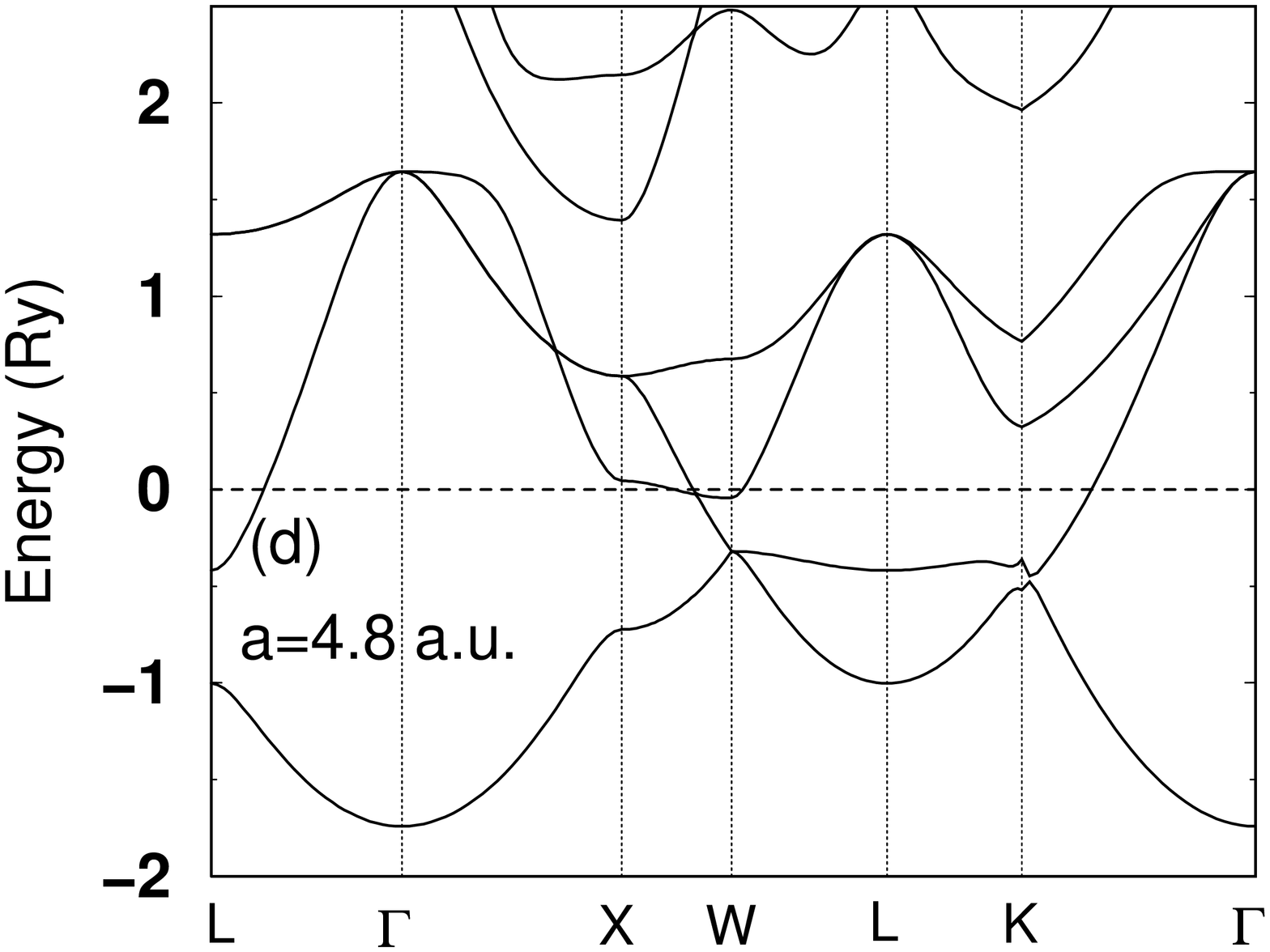}
\caption[]{
           FP-LMTO energy bands in fcc Fe for four different lattice parameters. The flat energy
 band between the symmetry points X and W, which  provides strong electron-phonon coupling, also renders
the structure unstable at higher volumes. The zero of energy has been set at the Fermi level.
}
\label{fig2}
\end{figure*}

In Fig.\ref{fig2} we show the FP-LMTO energy bands in fcc B as a function of the lattice parameter 'a'.
Calculations were carried out with a double-$\kappa$ $spd$ LMTO basis set for describing the valence bands. 
The charge densities and
potentials were represented by spherical harmonics with $l\leq$6 inside the nonoverlapping 
(touching) muffin-tin
spheres and by plane waves with energies $\leq$495 Ry in the interstitial region. Brillouin zone integrations
were carried out with the tetrahedron method\cite{tetra} using a 24,24,24 division of the 
Brillouin zone, corresponding
to 413 wave vectors in the irreducible part.
 The extraordinarily large band-widths in Fig.\ref{fig2}  are due to the extreme pressures of the systems.
The fcc phase is stable for a=4.4 a.u. and less, and unstable for a=4.6 a.u. and higher values. The
smallest volume per atom at which the bct phase is stable is around 3.55 \AA$^3$, according to 
the volume-energy curves
shown above. This volume corresponds to a fcc lattice parameter of $4.576$ a.u. One of the two 
energy bands crossing the
Fermi level between the two symmetry points X and W becomes very flat at larger volumes. 
The character of the flat energy band between X and W is primarily
 '$s$' and '$p$': at W it is purely of '$p$' and at X it is purely of  '$s$' character.
Papaconstantopoulos and Mehl\cite{papa} have speculated that this flat band is responsible for
superconductivity of the fcc phase in boron, pointing out that a similar flat band at the Fermi 
level appears
in MgB$_2$. Fermi surface plot of Papaconstantopoulos and Mehl\cite{papa} shows hole pockets near 
X and electron
pockets near W, as would also be expected according to the present calculations. 
 Although we have not verified this explicitly, it is very likely that the 
 flat band between X and W points contributes significantly to the electron-phonon coupling
 in fcc boron. 
 As this band becomes flatter  for larger lattice parameters,
the electron-phonon coupling increases.
However, this large electron-phonon coupling also makes the fcc phase unstable at 
higher volumes via softening of
X-point transverse phonons, and brings about the fcc$\rightarrow$ bct transition,  as discussed below.
 The steep band crossing the Fermi level between X and W is of '$p$' character.  

\subsection{Phonons and Electron-phonon Coupling}
\begin{figure*}
\includegraphics[angle=270,width=16.0cm]{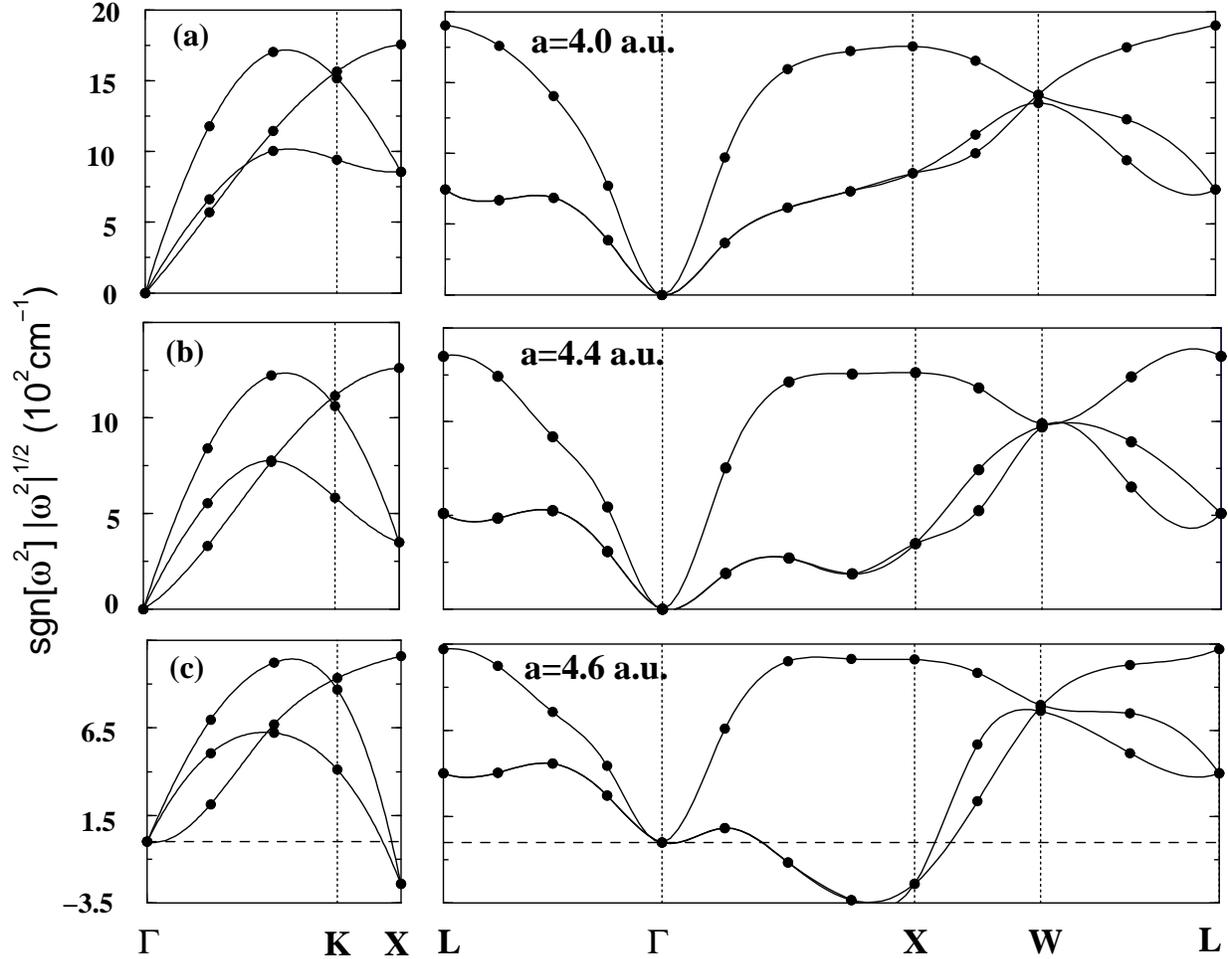}
\caption[]{Calculated phonon spectra for fcc B for the lattice parameters (a) $a$ = 4.0 a.u.,
(b) $a$ = 4.4 a.u. and (c) $a$ = 4.6 a.u.  The dots represent the calculated phonon frequencies, with
solid lines providing only a guide to the eye. The softening of the transverse phonons at the 
symmetry point X
occurs as the system expands to a volume above 3.4 \AA$^3$/atom, in accordance with the 
earlier results of 
Mailhoit \etal (Ref.\onlinecite{mcmahan}).}
\label{fig3}
\end{figure*}

\begin{figure}
\includegraphics[angle=270,width=12.0cm]{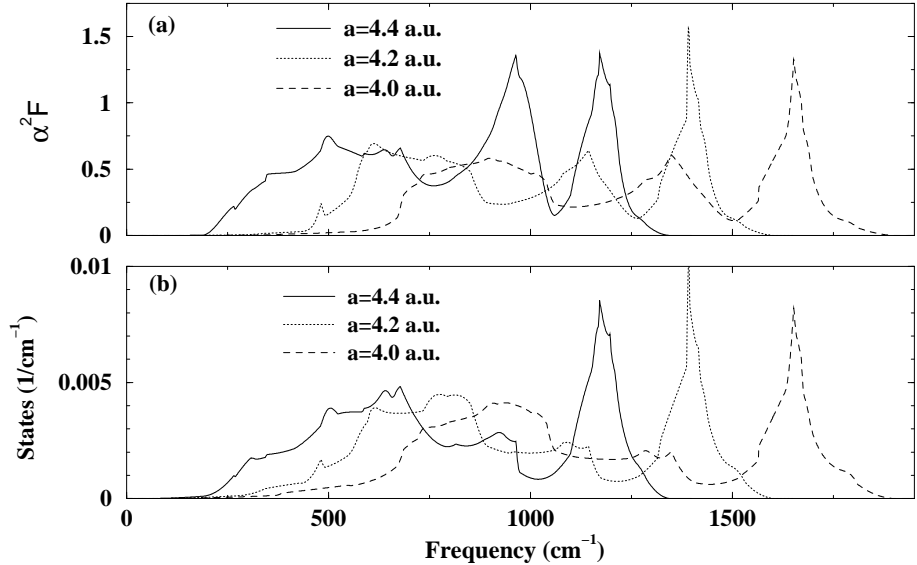}
\caption[]{ Eliashberg spectral function (a) and phonon density of states (b) for fcc B, for three
different lattice parameters.                                                                  
}
\label{fig4}
\end{figure}

In Fig.\ref{fig3} we show the variation of the  calculated phonon spectra with the lattice
parameter  for fcc B. The phonon frequencies were calculated for 29 wave vectors in the irreducible
part of the Brillouin zone (BZ), resulting from a 8,8,8 division of  the BZ. This division of the
BZ yields only a small number (3-4) of wave vectors along the symmetry directions. The solid lines
in Fig.\ref{fig3} should be taken only as  a guide to the eye, rather than actual phonon branches
with correct branch crossings.
 With increasing volume, the phonon frequencies soften throughout the Brillouin
zone, as expected. However, the softening of the  two transverse phonon branches near the X 
point is most pronounced. 
The X-point transverse phonons are related to the 180$^0$ out of phase vibration of  
the two atoms in a bct unit cell (c/a=
1.414 for fcc lattice). Softening of these phonons with increasing volume indicates growing 
instability  of the fcc structure with respect to the c/a ratio, and acts as a precursor
to the fcc$\rightarrow$bct phase transition. This phonon-softening has two important effects: it  
 increases the phonon density of states at low frequencies beyond the  usual parabolic DOS given by
the Debye (continuum) model and increases the electron-phonon coupling in the low frequency region
(partly due to increased phonon DOS and partly due to increased electron-phonon matrix element). 
 These effects are clearly seen in
 Fig.\ref{fig4}, where the calculated Eliashberg spectral function (upper panel (a)) and
the phonon density of states (lower panel (b)) are displayed for three different lattice parameters. 
 The sharp peaks in both the density of states and the Eliashberg 
spectral function at the high end of the spectrum are due to the lack of dispersion in the
 longitudinal phonon band close to the X point  in  the $\Gamma$-X direction. The lack of dispersion is
 enhanced  as the atoms move further apart, 
 giving a higher density of states and  consequently  higher electron-phonon coupling.

 In short, the transition from fcc to bct phase with increasing volume (lowering of pressure) 
 is driven by
 softening of transverse X-point  phonons, driving the system to a  c/a ratio different from 
 (in this case, lower than) the fcc value of 1.414.
 The system in this pressure/volume region shows strong electron-phonon coupling, 
 suggesting the possibility of
 superconducting phase with a relatively high transition temperature $T_c$.

\section {The bct Phase}
\subsection{Energy bands}

\begin{figure}
 \includegraphics[angle=270,width=8.0cm] {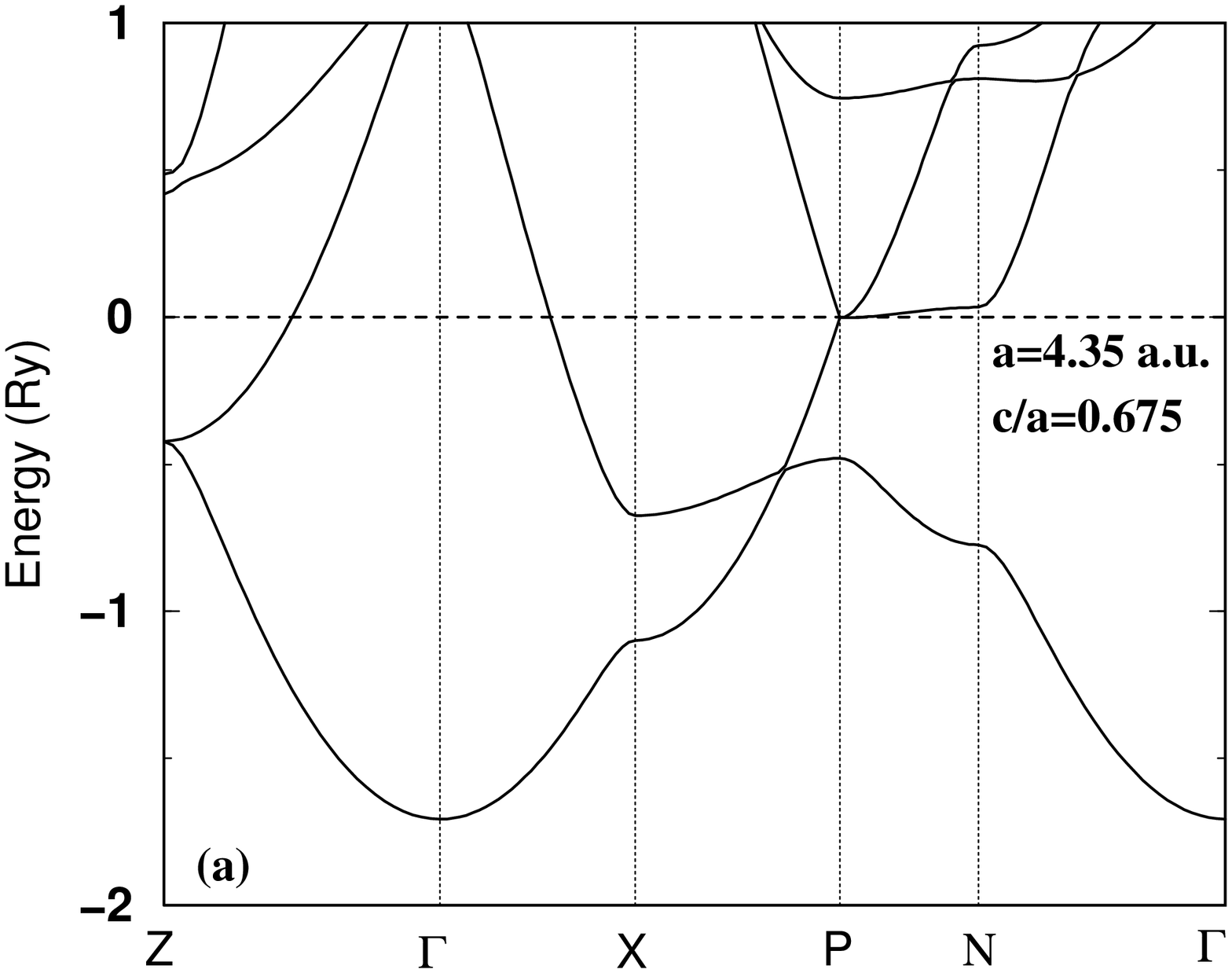}
 \includegraphics[angle=270,width=8.0cm] {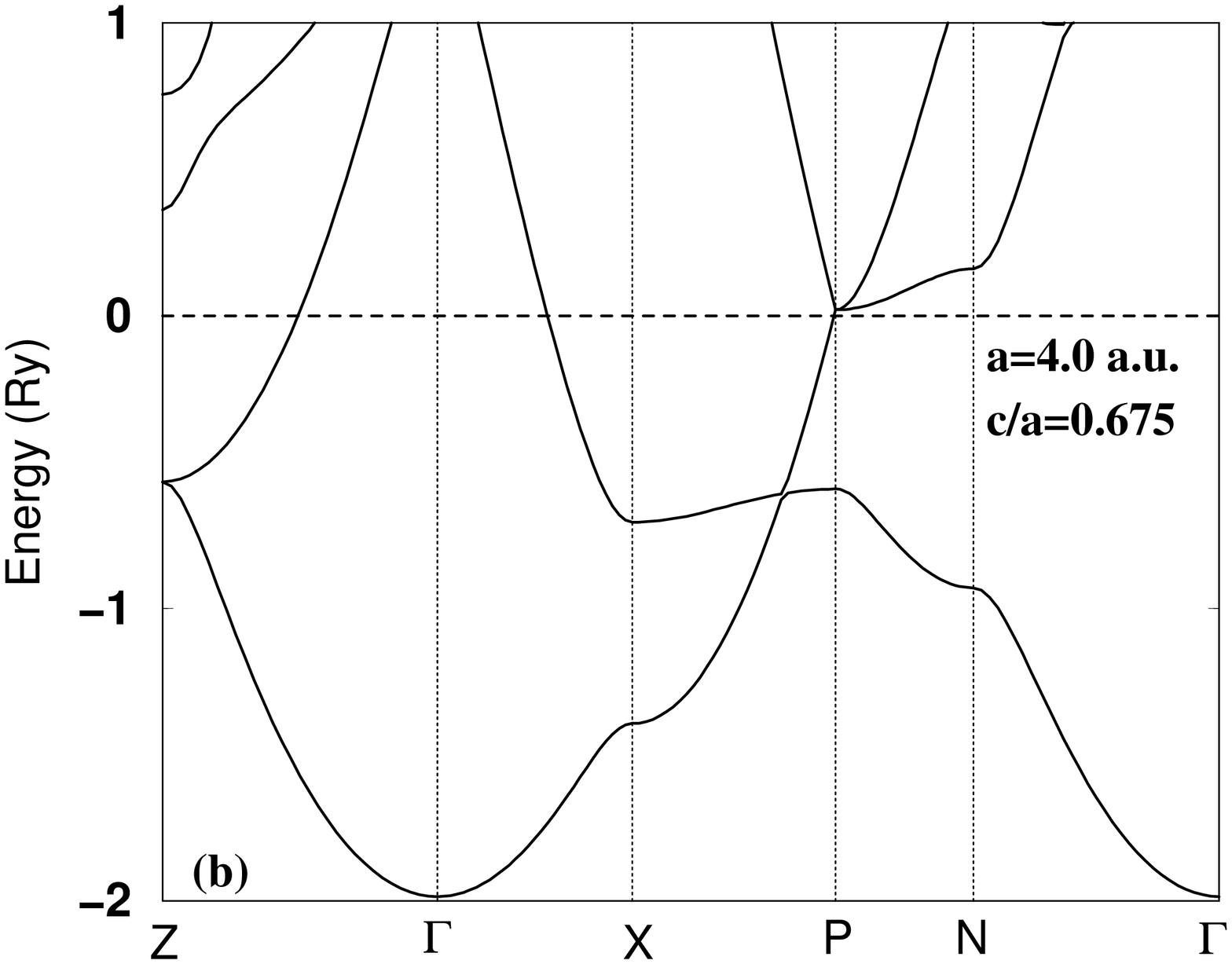}
 \includegraphics[angle=270,width=8.0cm] {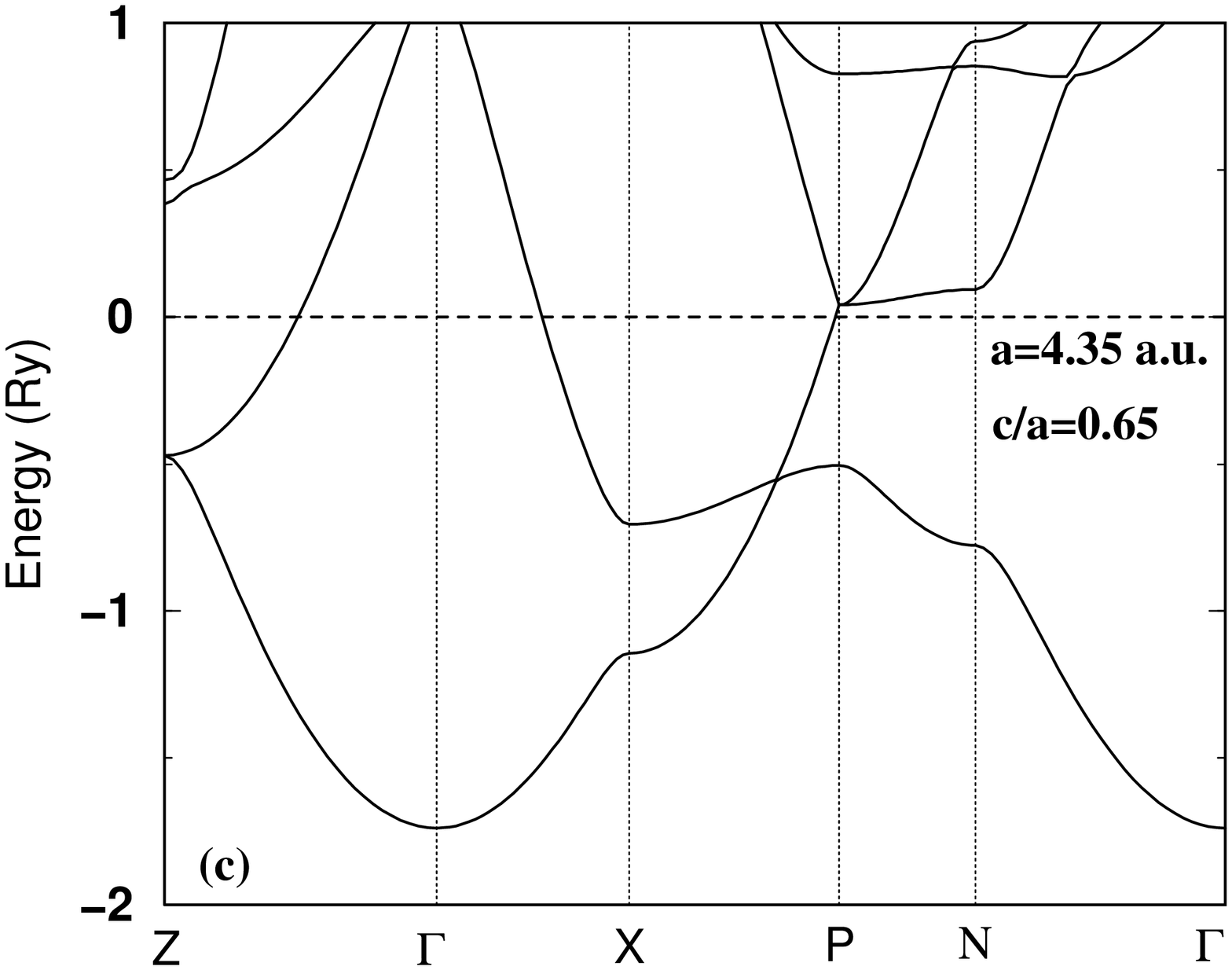}
 \caption[]{ FP-LMTO band structure of bct B for lattice parameter $a$=4.35 a.u., $c/a$=0.675 (a), 
  $a$=4.0 a.u., $c/a$=0.675 (b), and $a$=4.35 a.u., $c/a$= 0.65 (c). The zero of
  energy has been set at the Fermi level.}
 \label{fig5}
 \end{figure}

In Fig.\ref{fig5} we show the  FP-LMTO bands structure of bct B for $a$=4.35 a.u. 
for two different c/a ratios:
0.675 and 0.65. Another band structure for c/a=0.675, but $a$= 4.0 a.u. is also shown for comparison.
Calculations were carried out with a double-$\kappa$ $spd$ LMTO basis set for describing the valence bands. 
The charge densities and
potentials were represented by spherical harmonics with $l\leq$6 inside the nonoverlapping 
(touching) muffin-tin
spheres and by plane waves with energies $\leq$400 Ry in the interstitial region. Brillouin zone 
integrations
were carried out with the tetrahedron method\cite{tetra} using a 30,30,30 division of the 
Brillouin zone, corresponding
to 1992 wave vectors in the irreducible part.

As in the fcc case, the band widths are abnormally large due to extreme high pressures of the systems.
 The change in the band structure is understandable as a result of decreasing volume from (a) to (b). 
 In panel (c) of Fig.\ref{fig5} we show the bands for a c/a ratio of 0.65, for which one of the
 transverse  N-point phonons has imaginary frequency. There is no discernable feature in the band structure
 that would indicate the softening of these phonons. For small c/a ratio the boron chains along the 
 c-directions
 become unstable against transverse vibrations. This is also found to occur for the c/a ratio of 
 0.675 for lattice
 parameters in excess of $a$=4.35 a.u. At such large volumes boron presumably enters the nonmetallic
 icosahedral phase. The bands at the Fermi level are mostly of $p_x$ and $p_y$ character. 

\subsection{Phonons and Electron-phonon Coupling}
\begin{figure}
\includegraphics[angle=270,width=12.0cm]{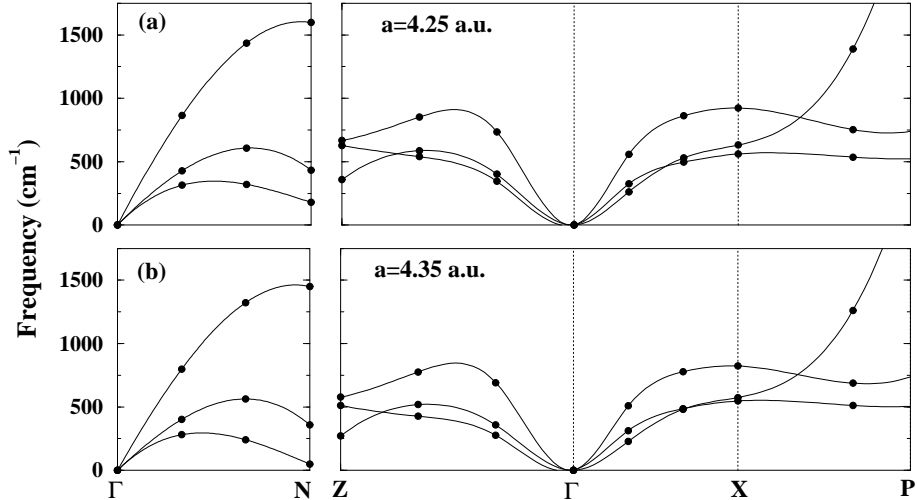}
\caption[]{
 Calculated phonon spectra for bct B for the lattice parameters (a) a=4.25 a.u. and (b) 
a=4.35 a.u. for the c/a ratio of 0.675. The dots represent the calculated       
frequencies, while the solid lines have been drawn through them to guide the eye.
For lattice parameters higher than 4.35 a.u. the structure becomes unstable 
with the N-point transverse phonons becoming soft first.} 
\label{fig6}
\end{figure}

In Fig.\ref{fig6} the phonon spectra of bct B for two different lattice parameters and with 
the c/a ratio of 0.675
are shown. The linear response calculations for the phonon properties were carried out for 
30 points in the
IBZ resulting from a 6,6,6 division of the BZ. The solid lines in Fig.\ref{fig6} 
have been drawn through the
calculated frequencies to provide a guide to the eye and should not be interpreted 
as the actual phonon branches.

With increasing values of the lattice
parameter phonons at other symmetry points become soft as well. 
 Thus our L-R calculations yield real phonon frequencies for a small range of volume 
 for the chosen c/a ratio
 of 0.675. For c/a=0.65 the N-point transverse phonons are found to be imaginary for all lattice
  parameters corresponding to volumes where the bct phase should be stable according to 
  the energy-volume curve
  in Fig.\ref{fig1}. Of all the phonons the ones at the N point are found to be most 
  sensitive to the c/a ratio. 

\begin{figure}
\includegraphics[angle=270,width=12.0cm] {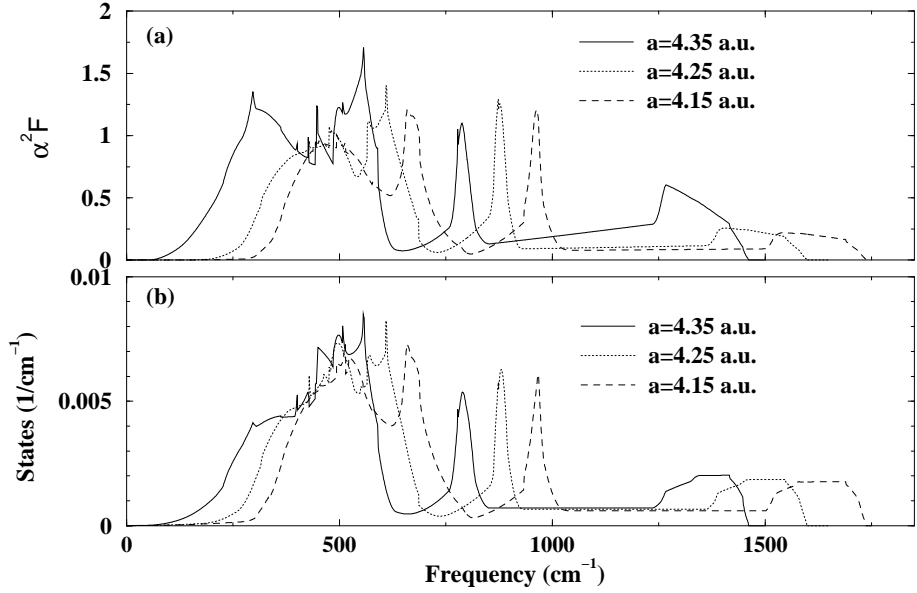}
\caption[]{Eliashberg spectral function (a) and phonon density of states (b) for bct B, for three
different lattice parameters, and c/a=0.675.
}
\label{fig7}
\end{figure}

In Fig.\ref{fig7} the phonon density of states and the Eliashberg spectral functions
for three different lattice parameters are shown. Unlike in the fcc phase the bulk of the electron-phonon
coupling is via lower frequency phonons, as most of the spectral weight in Fig. 7(b) comes from the lower
half of the allowed frequency range. As in the fcc case, the Eliashberg spectral function and the
phonon density of states follow each other closely. There is no disproportionately large
contribution from a particular mode.

\section{Superconducting Transition Temperatures and Summary of Results}

Superconducting transition temperatures $T_c$ were calculated by solving the isotropic Eliashberg equation. 
We used the calculated Eliashberg spectral function
along with the
following procedure to determine the Coulomb pseudopotential $\mu ^{\ast }(\omega
_{c}).$ We start by assuming $\mu(E_F)=1.0$. This value is consistent with that for MgB$_2$,
where  the measured $T_c \sim$ 40 K can be reproduced with the linear response results of phonon frequencies
and Eliashberg function with $\mu \sim 0.85$. $\mu ^{\ast }(\omega_{c})$ was calculated from
  \begin{equation*}
  \mu ^{\ast }(\omega _{c})=\frac{\mu(E_F)}{1+\mu(E_F)\ln (E_F/\omega _{c})}\;.
  \end{equation*}

   Thus from the calculated Fermi energies $E_F$  we obtain
   $\mu ^{\ast }$ for all volumes, with the cut-off frequency $\omega_{c}$ assumed to be
   ten times the maximum phonon frequency. A different value of $\mu(E_F)$, e.g. 0.8 or 0.7, would
   result in a minor change in the calculated $T_c$ and would have no effect on the nature of variation
   of $T_c$ with volume or pressure.

   The results of our calculation are summarized in Table \ref{table1}, where, for
   comparison, results for MgB$_2$ and two typical low $T_c$ superconductors
   fcc Pb and bcc Nb obtained via the same method are also presented. 
    Both for fcc and bct boron the Hopfield parameter $\eta$ increases with decreasing volume per atom.
    This trend is also supported by the rigid muffin-tin approximation results of 
     Papaconstantopoulos and Mehl\cite{papa}. The Hopfield parameter in these high pressure phases
     of boron are 2-4 times larger than in MgB$_2$ and about 10-15 times larger than in typical low $T_c$
     metals becauses of the vastly reduced distances between the atoms.
      In fcc boron $\eta$ is larger than in the bct phase, since the fcc phase is the equilibrium phase
     at higher pressures or lower volumes. Phonon frequencies  $\omega$ 
     in the high pressure bct and fcc phases of 
     boron are much higher than
     in MgB$_2$ or a typical metal. As a result the increase in the electron-phonon coupling parameter
      $\lambda$, which is proportional to $\eta/\langle\omega^2\rangle$  is less drastic (than   
      in $\eta$) when compared with  MgB$_2$ or a low $T_c$ superconductor
      like Pb or Nb. The area under the Eliashberg spectral function vs. $\omega$ curve $A$ is 5-15
      times larger for fcc or bct boron than for a typical electron-phonon superconductor, because
      the coupling of electrons in these high pressure phases takes place via very high frequency phonons.

\begin{figure}
	    \includegraphics[angle=270,width=12cm] {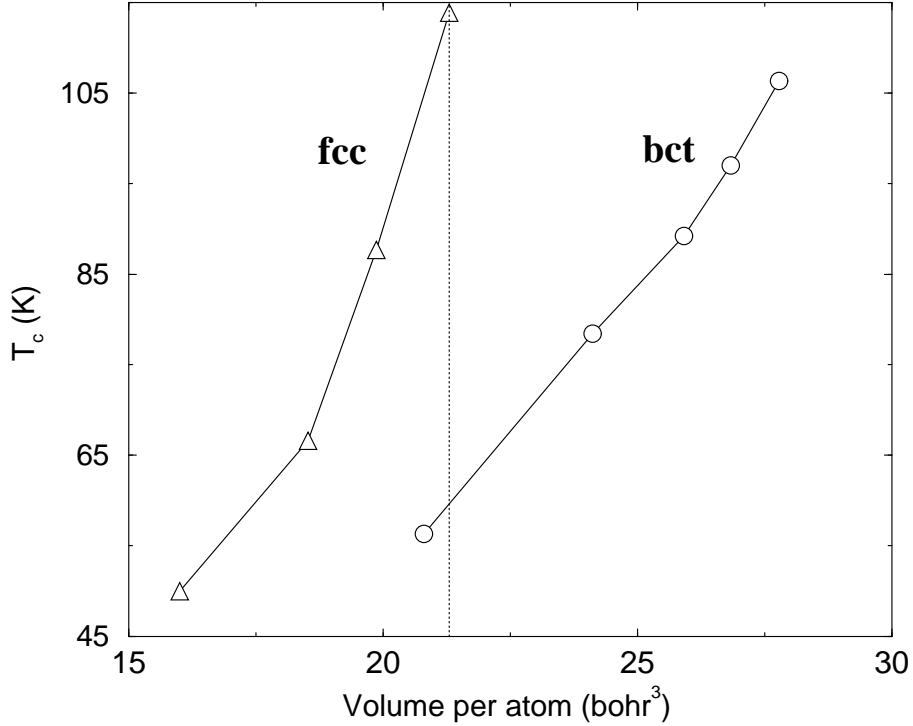}
	       \caption[]{Variation of the superconducting transition temperature
		  $T_c$ with volume per atom. The vertical dashed line is drawn at volume per atom equal to
			21.3 bohr$^3$, estimated to be the dividing line between the bct and the fcc phases.}
			      \label{fig8}
				    \end{figure}

The variation of the superconducting transition temperature
   $T_c$ with volume per atom is shown in Fig.\ref{fig8}, where the vertical dashed line is drawn at 
   volume per atom equal to
      21.3 bohr$^3$, estimated to be the dividing line between the bct and the fcc phase.

      For both bct and fcc boron the phonon frequencies rise faster with increasing pressure than the
      Hopfield parameter, resulting in a decrease in electron-phonon coupling and $T_c$. In the
      study  of the fcc phase by Papaconstantopoulos and Mehl\cite{papa} no calculation of phonon 
      spectra as a function 
      of pressure was done, and the authors speculated that if the phonon frequencies do not rise as fast
      as the Hopfield parameter, $T_c$ should rise with increasing pressure. The present calculations
      reveal two important results: $T_c$ decreases with increasing pressure despite rise in the Hopfield
      parameter and that the calculated $T_c$'s are in the range 50-100 K, much higher than the measured
      values. In the fcc phase electron-phonon coupling rises strongly due to softening of X-point 
      transverse phonons as volume per atom approaches the fcc$\rightarrow$bct phase transition. Similarly
      in the bct phase electron-phonon coupling is very high near the bct$\rightarrow$icosahedral phase
      change due to softening of N-point transverse phonons. Unlike in MgB$_2$ there is no disproportionately
       high contribution to electron-phonon coupling from a particular phonon mode. The Eliashberg spectral
       function follows faithfully the phonon density of states.
   
   The striking differences between the results of this theoretical investigation and the experimental results
   beg for a suitable explanation. While we cannot rule out the possibility that the electron-phonon coupling
   and therefore $T_c$ is somehow overestimated in the calculation, it would be hard to deny the validity of
   the results altogether. The values of the Hopfield parameter and the trend in their variation with pressure 
   are similar to
   those found via a less rigorous rigid muffin-tin approximation calculation of 
   Papaconstantopoulos and Mehl\cite{papa}.
   The phonon calculations correctly predict the fcc$\rightarrow$bct transition and also 
   the onset of instability 
   of the bct phase (presumably against the icosahedral phase) as the pressure is lowered or 
   the volume per atom
   is increased. Since the systems studied do not represent cases of extreme electron-phonon coupling
    (see e.g., J. M. An {\it et al.}\cite{libc}) via a
   single mode or a few modes, superconductivity in these  systems does not seem to be special.
    
    On the experimental front, the possibility that the measurements were not carried out 
    for a single (pure)
    phase cannot be ruled out, as was pointed out by the authors themselves \cite{science}. 
    It is possible that
    the systems studied were in mixed phases with co-existing icosahedral and bct clusters 
    and the superconductivity
    of the sample was due to the proximity effect of the bct clusters. This would explain 
    why the measured $T_c$
    was so much lower than what would be expected for the pure bct phase. It is also 
    possible that with increasing
    pressure the samples were just evolving towards pure bct phase, which would explain 
    the observed increase in
    $T_c$ with pressure. The results of the present work strongly suggest that the experimental 
    work on high
    pressure phases of boron should be repeated, with particular emphasis on the 
    phase of the sample. It is 
    expected that if the phase is bct or fcc (or a mixture of the two), superconducting 
    transition temperatures 
    should be much higher than previously reported.

   \begin{table*}
   \caption{ Variation of superconducting transition temperature $T_c$ with volume in the bct and fcc phases
   of boron.
    $V$ = volume per atom, $\omega$ stands for phonon frequency and $\langle \cdots\rangle$s denote averages, 
    $\bar{\omega}=\sqrt{\langle\omega^2\rangle}$. $\omega_m$ is the maximum phonon frequency, 
    $\eta$ is the Hopfield parameter and  
    A denotes the area under the phonon frequency versus  Eliashberg spectral function curve $\alpha^2F$.
    For comparison, results for hexagonal MgB$_2$  and two typical low $T_c$ superconductors, fcc Pb 
    and bcc Nb, are also presented. $a_0$ denotes the bohr radius. Results for MgB$_2$ 
    are taken from Reference \onlinecite{kong}}
   \label{table1}
   \hrule
   {\bf bct Boron}\\
   \begin{ruledtabular}
   \begin{tabular}{lccccccccc}
   $V$($a_0^3$)&$\omega_m$(meV)& $\bar{\omega}$(meV)&$\langle\omega\rangle$(meV)&$\mu ^{\ast }(\omega_{c})$
   &$\eta$(Ry/$a_0^2)$&$\lambda$&A(meV)&$T_c$(K)\\
   20.80  & 254 & 153& 97  & 0.297 & 0.624&1.025 & 50.06 & 56.33\\
   24.12  & 215 & 123& 77  & 0.289 &0.552   & 1.396 & 55.00 & 78.44 \\
   25.91  & 198 & 110& 68  & 0.285 & 0.549&1.702 & 60.18 & 89.24\\
   26.83  & 189 & 70 & 60.4 & 0.283 & 0.562 & 2.016 & 64.95 & 97.02\\
   27.78 &  181 & 76.8 & 42  & 0.281 &0.519& 2.735 & 71.99 & 106.3 \\
   \end{tabular}
   \hrule
   {\bf fcc Boron}\\
   \begin{tabular}{lccccccccc}
   $V$($a_0^3$)&$\omega_m$(meV)&$\bar{\omega}$(meV)&$\langle\omega\rangle$(meV)&$\mu^{\ast }(\omega_{c})$
   & $\eta$(Ry/$a_0^2)$& $\lambda$&A(meV)&$T_c$(K)\\
   16 & 236 & 143& 136  & 0.277 &0.926& 0.842 & 57.52 & 50.00\\
   18.52 & 199 & 116& 108 & 0.269 &0.752& 1.039 & 57.28 & 66.62\\
   19.87 & 183 & 102&94.4  & 0.266 &0.744& 1.305 & 63.30 & 87.70\\
   21.30 & 167 & 88&77.6  & 0.262 &0.797& 1.787 & 75.12 & 113.9\\
   \end{tabular}
   \hrule
   {\bf MgB$_2$} (Reference \onlinecite{kong})\\
   \begin{tabular}{lccccccc}
   $\omega_m$(meV)&$\bar{\omega}$(meV)&$\langle\omega\rangle$(meV)&$\mu^{\ast }(\omega_{c})$&$\eta
   $(Ry/$a_0^2)$&$\lambda$&A(meV)&$T_c$(K)\\
   100 & 66.6 & 64.1 & 0.14 & 0.210 & 0.867 &27.77 & 40
   \end{tabular}
   \hrule
      {\bf fcc Pb}\\
      \begin{tabular}{lccccccc}
      $\omega_m$(meV)&$\bar{\omega}$(meV)&$\langle\omega\rangle$(meV)&$\mu^{\ast }(\omega_{c})$
      &$\eta$(Ry/$a_0^2)$&$\lambda$&A(meV)&$T_c$(K)\\
      9.70 & 8.96 & 5.77 & 0.12 & 0.051 & 1.27 & 3.76 & 6.77
      \end{tabular}
      \hrule
      {\bf bcc Nb}\\
      \begin{tabular}{lccccccc}
      $\omega_m$(meV)&$\bar{\omega}$(meV)&$\langle\omega\rangle$(meV)&$\mu ^{\ast }(\omega_{c})$
      &$\eta$(Ry/$a_0^2)$&$\lambda$&A(meV)&$T_c$(K)\\
      27.5 & 24.1 & 15.6 & 0.45 & 0.179 & 1.33 & 10.82 & 11.76
      \end{tabular}
   \end{ruledtabular}
   \end{table*}

\begin{center}
ACKNOWLEDGMENTS
\end{center}
SKB would like to thank Jens Kortus for bringing the experimental results on Boron to his attention.
TK would like to acknowledge financial support from the Institute of Fundamental Chemistry, 
Kyoto, Japan. Partial financial support
for this work was provided by Natural Sciences and Engineering Research Council 
of Canada.

\end{document}